\documentclass{aa}
\usepackage{graphicx}
\usepackage{txfonts}
\usepackage{natbib}

\begin{document}
  \title{Two new candidate ultra-compact X-ray binaries}

  \author{C. G. Bassa\inst{1}
    \and P. G. Jonker\inst{2,3,1}
    \and J. J. M. in 't Zand\inst{2,1}
    \and F. Verbunt\inst{1}}

  \institute{
    Astronomical Institute, Utrecht University, PO Box 80\,000, 3508
    TA Utrecht, The Netherlands\\ \email{c.g.bassa@astro.uu.nl}
    \and SRON Netherlands Institute for Space Research, Sorbonnelaan
    2, 3584 CA Utrecht, The Netherlands
    \and Harvard-Smithsonian Center for Astrophysics, 60 Garden
    Street, MS83, Cambridge, Massachusetts, USA
  }
  
  \date{Received / Accepted}
  
  \abstract{We present the identification of the optical counterparts
    to the low-mass X-ray binaries 1A~1246$-$588 and 4U~1812$-$12. We
    determine the X-ray position of 1A~1246$-$588 from
    \emph{ROSAT}/PSPC observations and find within the error circle a
    blue star with $V=19.45$, $B-V=0.22$ and $R-I=0.22$ which we
    identify as the counterpart. Within the \emph{Chandra} error
    circle of 4U~1812$-$12, a single star is present which appears
    blue with respect to the stars in the vicinity. It has $R=22.15$,
    $R-I=1.53$. Distance estimates for both systems indicate that the
    optical counterparts are intrinsically faint, suggesting that they
    are ultra-compact X-ray binaries. These identifications would
    increase the number of candidate ultra-compact X-ray binaries from
    2 to 4, whereas orbital periods are measured for only 7 systems in
    the Galactic disk.
    
     } 
  \maketitle

  \section{Introduction}
  The canonical low-mass X-ray binary consists of a neutron star or
  black hole and a low-mass main-sequence or (sub)giant donor star,
  and has an orbital period longer than one hour, up to several
  hundred days.

  Recently, it has been found that the class of ultra-compact low-mass
  X-ray binaries makes up about half (5 out of 12) of the low-mass
  X-ray binaries in globular clusters (e.g., review by
  \citealt{vl04}), whereas a growing number of such systems is also
  discovered in the Galactic disk (7 with measured periods, and 2
  candidates, see e.g., \citealt{jc03,njmk04,wc04}). In addition,
  observations with the Wide Field Cameras (WFCs) of \emph{BeppoSAX}
  have found a new class of low-mass X-ray binaries, bursters with
  (very) low persistent X-ray emission \citep{cbn+01}. The
  distribution of the sources in this class is more concentrated
  towards the Galactic center than that of the canonical low-mass
  X-ray binaries \citep{cvz+02}.

  To elucidate the evolutionary status and history of these systems,
  observations at longer wavelengths, in particular optical/infrared,
  are crucial. Such observations may reveal the orbital period,
  directly or indirectly (e.g., \citealt{vpm94}), or provide
  information about the donor and its chemical composition (e.g.\
  \citealt{njmk04}). The first step to such studies is the optical
  identification of the X-ray binary. In this letter we report two new
  optical identifications, which may be ultra-compact X-ray binaries.
  
  \section{X-ray Observations}\label{sec:xrayobservations}
  4U~1812$-$12 ($l=18\fdg03$, $b=2\fdg40$) has been observed by
  various X-ray satellites, but the observations with
  \emph{BeppoSAX}/WFC and \emph{Chandra} are most relevant for this
  paper. Type I X-ray bursts with photospheric radius expansion have
  been observed with the former, providing an unabsorbed bolometric
  peak flux of $(1.5\pm0.3)\times10^{-7}$\,erg\,cm$^{-2}$\,s$^{-1}$
  \citep{cbn+00}. A 1\,ks observation with the back-illuminated S3 CCD
  aboard \emph{Chandra} was analyzed by \citet{wpk+03}, yielding an
  accurate position of 4U~1812$-$12;
  $\alpha_\mathrm{J2000}=18^\mathrm{h}15^\mathrm{m}06\fs18$,
  $\delta_\mathrm{J2000}=-12\degr05\arcmin47\farcs1$, with an
  uncertainty limited by the \emph{Chandra} bore-sight ($0\farcs6$,
  90\% confidence; \citealt{akc+00}). A flux of
  $4.4\times10^{-10}$\,erg\,cm$^{-2}$\,s$^{-1}$ (1--10\,keV) and
  absorption of $N_\mathrm{H}=(1.1\pm0.2)\times10^{22}$\,cm$^{-2}$
  were determined from spectral fits.

  \begin{figure*}
    \centering
    \includegraphics[width=85mm]{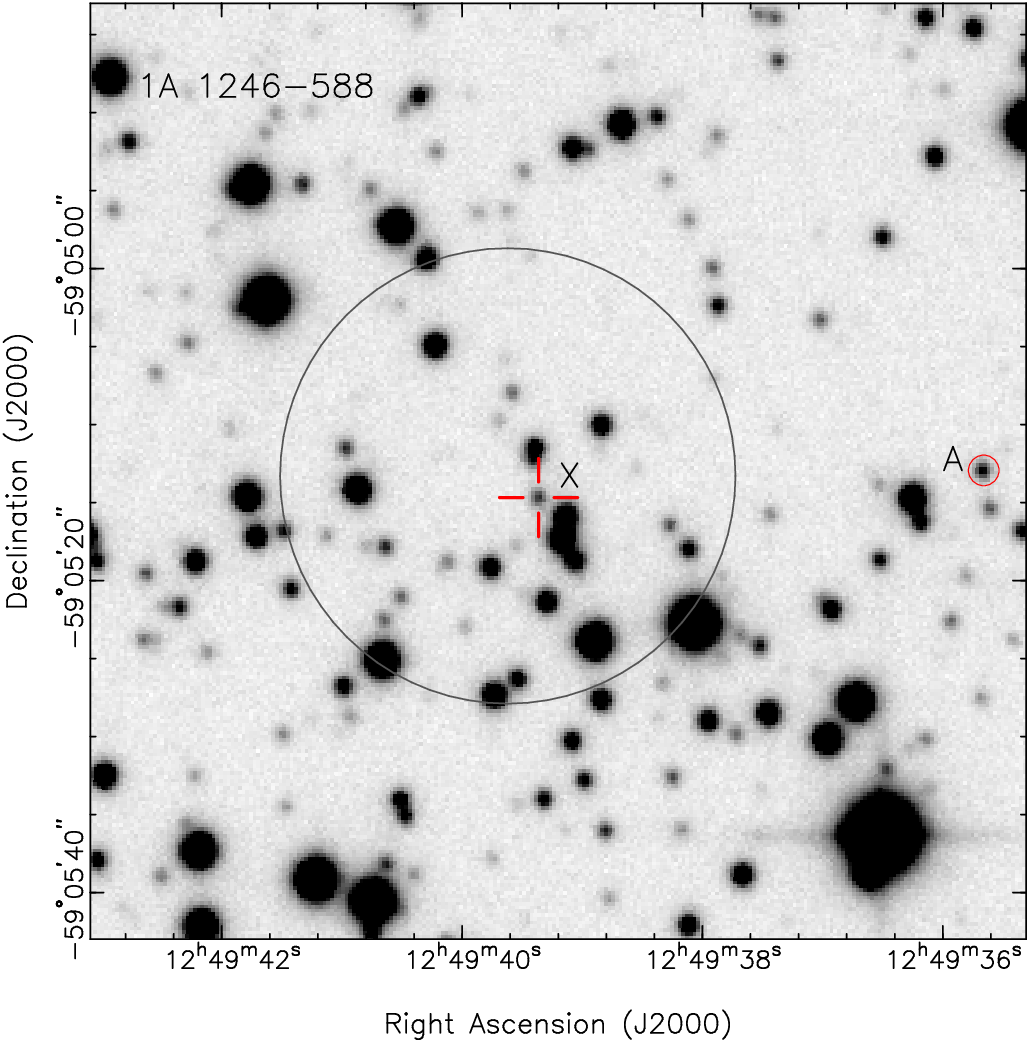}
    \includegraphics[width=85mm]{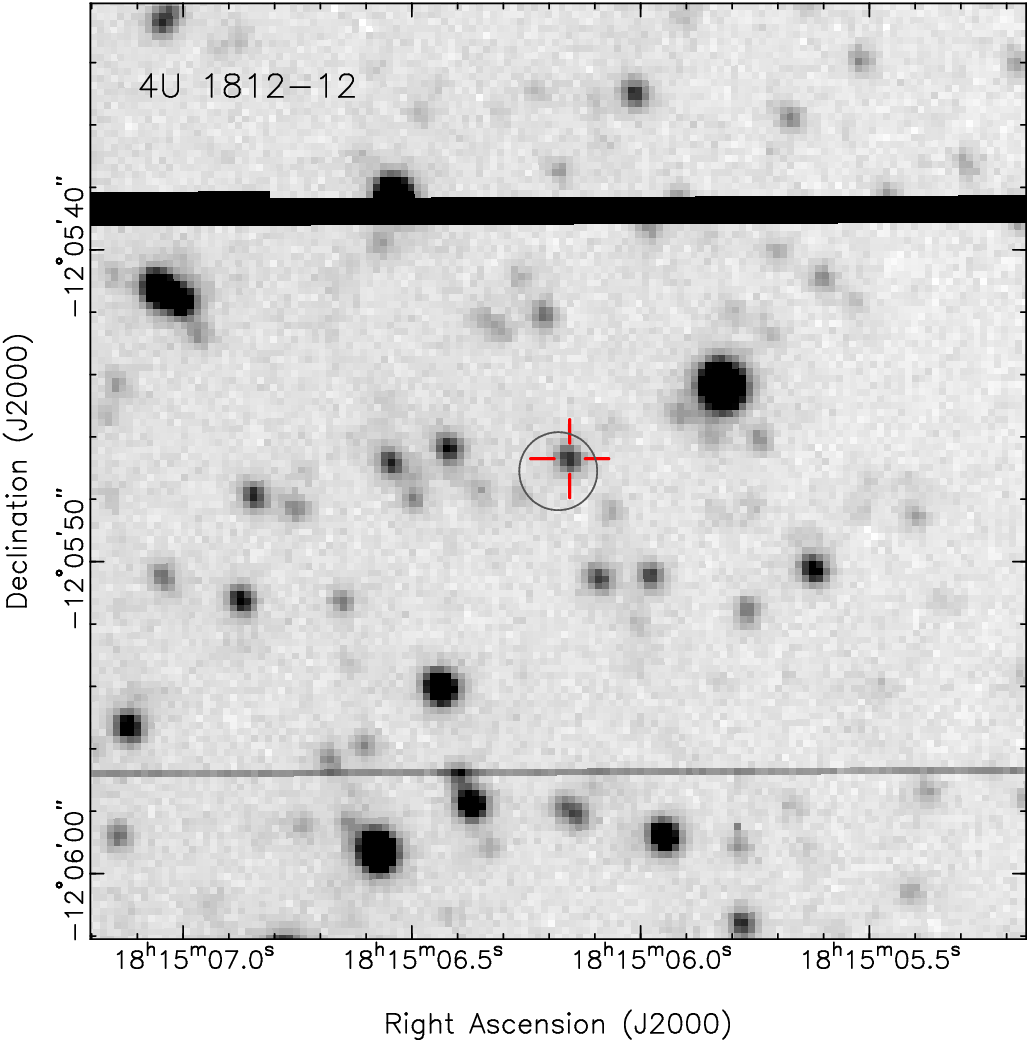}
    \caption{(\emph{left}) A $1\arcmin\times1\arcmin$ subsection of
    the combined 2\,min $R$-band WFI image of 1A~1246$-$588. The 95\%
    confidence error circle on the \emph{ROSAT} position has a radius
    of $14\farcs6$ and is depicted with the large circle. The proposed
    optical counterpart, star~X, is indicated with the tick marks,
    while another blue star, star~A, which is located outside the
    error circle, is encircled. (\emph{right}) A
    $30\arcsec\times30\arcsec$ subsection of the 5\,min $R$-band IMACS
    image of 4U~1812$-$12. The horizontal black bar in the top part of
    the image is a read out streak of a bright star, whereas the grey
    streak in the bottom part is a dead column of the CCD. The
    \emph{Chandra} error circle (95\% confidence, $1\farcs25$ in
    radius) is depicted with the circle. The proposed optical
    counterpart is indicated with the tick marks.  }
    \label{fig:fc}
  \end{figure*}

  1A~1246$-$588 ($l=302\fdg70$, $b=3\fdg78$) has received much less
  attention, though it has been observed serendipitously by several
  X-ray observatories. A type I X-ray burst was observed with
  \emph{BeppoSAX}/WFC \citep{phj+97} and a short (0.85\,ks) follow-up
  observation with \emph{ROSAT}/PSPC linked it to 1A~1246$-$588
  \citep{bhv+97}. In this observation $N_\mathrm{H}$ was measured to
  be $(2.9\pm0.9)\times10^{21}$\,cm$^{-2}$ and the 0.1--2.4\,keV flux
  was $1.7\times10^{-10}$\,erg\,cm$^{-2}$\,s$^{-1}$ \citep{bhv+97}.
  We have reanalyzed the \emph{ROSAT}/PSPC observations of the field
  of 1A~1246$-$588 using standard routines from the EXSAS distribution
  \citep{zbb+96}. In the 0.85\,ks exposure obtained in February 1997,
  the X-ray binary is only $7\arcmin$ off-axis, compared to
  $39\arcmin$ in the much longer PSPC observation from 1993, and the
  shorter observation provides the best source position. In this
  observation the X-ray binary has
  $\alpha_\mathrm{J2000}=12^\mathrm{h}49^\mathrm{m}39\fs61$,
  $\delta_\mathrm{J2000}=-59\degr05\arcmin13\farcs3$, with an internal
  uncertainty of $0\farcs3$ on each coordinate. The external
  uncertainty on this position, the uncertainty in the bore-sight of
  the satellite, is about $6\arcsec$ \citep{ayr04}. Due to the short
  exposure only a few X-ray sources are present in the PSPC
  observation, and none of them are coincident with bright
  stars. Thus, no bore-sight correction is possible and the
  uncertainty on the position is dominated by the pointing uncertainty
  of \emph{ROSAT}.

  1A~1246$-$588 and 4U~1812$-$12 have been persistently detected in
  X-rays by the All Sky Monitor (ASM) onboard the Rossi X-ray Timing
  Explorer (RXTE). Both during, a few days before and a few days after
  the time of the optical observations described below both sources
  were detected at daily averaged count rates of at least 1\,ASM count
  per second.

  \section{Optical Observations}\label{sec:opticalobservations}
  We retrieved archival observations of 1A~1246$-$588 obtained with
  the Wide Field Imager (WFI) at the ESO 2.2\,metre telescope on La
  Silla on March 26/27, 2000. A series of dithered 4 $B$, 6 $V$, 5 $R$
  and 5 $I$-band images were taken, all with exposure times of 2\,min
  under clear conditions with $0\farcs8$--$1\farcs0$ seeing. The field
  of 4U~1812$-$12 was imaged with the 6.5\,metre Inamori Magellan
  Areal Camera and Spectrograph (IMACS) at the Magellan Baade
  telescope on Las Campanas on July 6/7, 2005. A single 5\,min image
  was obtained in both $R$ and $I$ under $0\farcs6$ seeing. Both IMACS
  and WFI are mosaics of eight 4k$\times$2k detectors and we analyzed
  the images from the detector containing the X-ray binaries (chip 8
  for 4U~1812$-$12 and chip 2 for 1A~1246$-$588). The WFI images have
  a pixel scale of $0\farcs24$\,pix$^{-1}$, while IMACS observations
  were taken with $2\times2$ binning, yielding a pixel scale of
  $0\farcs22$\,pix$^{-1}$. The science images were corrected for bias
  and flatfielded with domeflats using standard routines running
  within MIDAS.

  Large-scale variations in the background of the $I$-band images,
  known as fringing, were present in the WFI observations. We
  corrected for this with a fringe frame. This fringe frame was
  constructed by median combining a set of 20 $I$-band images obtained
  earlier that night, such that it contained only the contributions of
  the sky and the fringe variations. The level of the sky was
  estimated and subtracted from this image, leaving only the fringe
  variations. This result was scaled to the fringe variations in the
  $I$-band images of the X-ray binary and subsequently subtracted from
  these images.

  For the WFI observations, the images taken through the same filter
  were aligned using integer pixel offsets and median combined to
  remove image artifacts and increase the overall signal-to-noise
  ratio. Finally, a $4\arcmin\times4\arcmin$ subsection of the
  averaged images, centered on the nominal position of the X-ray
  binary, was extracted and used for the astrometry and
  photometry. For 4U~1812$-$12, a $3\farcm8\times3\farcm8$ subsection
  of the IMACS $R$ and $I$-band images was extracted.

  For the astrometry of the WFI observations of 1A~1246$-$588 we
  measured the centroids of all 31 astrometric standards from the
  second version of the USNO CCD Astrograph Catalog (UCAC2;
  \citealt{zuz+04}) that overlapped with the $4\arcmin\times4\arcmin$
  subsection of the combined $R$-band image and that were not
  saturated and appeared stellar and unblended.  We removed one
  outlier that had a total residual of $0\farcs74$, and the remaining
  stars were used to compute an astrometric solution, fitting for
  zero-point position, scale and position angle. The astrometric
  solution has root-mean-square (rms) residuals of $0\farcs054$ in
  right ascension and $0\farcs047$ in declination. We used a similar
  approach for the IMACS observations of 4U~1812$-$12, but had to use
  the USNO-B1 catalog \citep{mlc+03} as very few UCAC2 standards
  overlapped with the IMACS images. About 70 USNO-B1 standards were
  used to calibrate a $3\farcm8\times3\farcm8$ subsection of the
  $R$-band image, giving a solution with rms residuals of $0\farcs19$
  in right ascension and $0\farcs21$ in declination.

  \begin{figure}
    \centerline{\resizebox{\hsize}{!}{\includegraphics{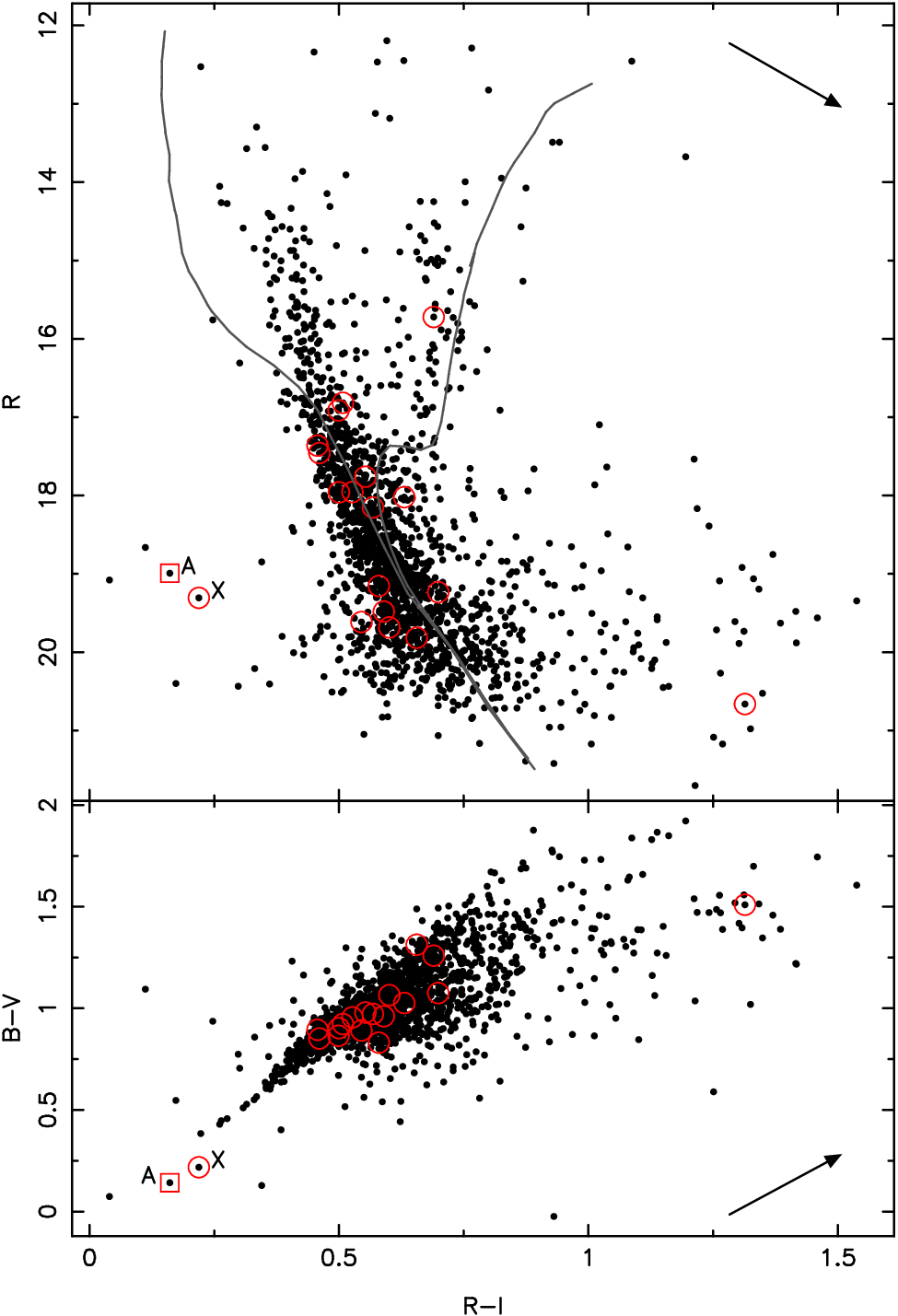}}}
    \caption{(\emph{top}) A colour-magnitude and colour-colour diagram
    (\emph{bottom}) of the $BVRI$ photometry of the 1A~1246$-$588
    region. Stars inside the \emph{ROSAT} error circle
    (Fig.~\ref{fig:fc}) are encircled. Also shown in the CMD are two
    solar metallicity isochrones from \citet{gbbc00}, placed at a
    distance of 4.0\,kpc and an absorption of $A_V=1.0$. The left is
    for an age of 0.1\,Gyr, while the right has an age of
    10\,Gyr. These isochrones are merely to guide the eye, as there
    will be a spread in distance, age, metallicity and absorption. The
    arrows indicate the effects of absorption, where the length of the
    arrow is for an extra absorption of $\Delta A_V=1.0$. Of the stars
    present in the error circle, star~X is exceptionally blue and the
    likely optical counterpart to 1A~1246$-$588. Star~A has similar
    colours and magnitudes as star~X, but is located outside the 95\%
    confidence error circle.}
    \label{fig:cmd1246}
  \end{figure}

  The DAOPHOT II package \citep{ste87}, running inside MIDAS, was used
  to determine instrumental magnitudes through point spread function
  (PSF) fitting. Aperture photometry of several bright stars was used
  to determine aperture corrections. For the calibration of the WFI
  observations we determined instrumental magnitudes of some 100
  photometric standards in the standard field SA\,98 and calibrated
  these against the calibrated values by \citet{ste00}, fitting for
  zero-point and colour coefficients. We assumed extinction
  coefficients of 0.22, 0.19, 0.14 and 0.11\,mag per airmass for $B$,
  $V$, $R$ and $I$-band, respectively, taken from the WFI
  webpage\footnote{http://www.ls.eso.org/lasilla/sciops/2p2/E2p2M/WFI/zeropoints/}.
  The rms residuals of the calibration were 0.04\,mag in $B$ and $R$,
  0.03 in $V$ and 0.05\,mag in $I$. The IMACS observations were
  calibrated using 11 standards in the T~Phe field, again using values
  from \citet{ste00} and fitting for zero-point and colour
  coefficients. The standard field was imaged at similar airmass as
  the 4U~1812$-$12 field and no extinction coefficients were used. The
  rms residuals of the calibration were 0.05\,mag in $R$ and 0.07\,mag
  in $I$.

  In Fig.~\ref{fig:fc} we present finding charts for the regions of
  1A~1246$-$588 and 4U~1812$-$12. We searched for optical counterparts
  to the X-ray sources in 95\% confidence error circles on the
  \emph{ROSAT} and \emph{Chandra} position. For 4U~1812$-$12, a single
  star is located in the $1\farcs25$ error circle, while several stars
  lie inside the $14\farcs6$ error circle on the position of
  1A~1246$-$588. We note that as both X-ray binaries are expected to
  have an accretion disk, the optical colours should display an excess
  of emission at blue wavelengths and thus appear blue, typically
  having $(B-V)_0\approx0.0$ \citep{vpm95}.

  In Figure~\ref{fig:cmd1246} we show the colour-magnitude diagram
  (CMD) and a colour-colour diagram of the $BVRI$ photometry of all
  stars on the $4\arcmin\times4\arcmin$ image of 1A~1246$-$588. To
  illustrate the structure seen in the CMD, we have overplotted
  isochrones from \citet{gbbc00}. There are 3 stars that are
  exceptionally blue (having $B-V<0.25$ and $R-I<0.25$) and two of
  them, stars A and X, are near the error circle
  (Fig.~\ref{fig:fc}). Star X has $V=19.45\pm0.02$, $B-V=0.22\pm0.03$,
  $V-R=0.14\pm0.03$ and $R-I=0.22\pm0.06$, while star A has
  $V=18.99\pm0.01$, $B-V=0.15\pm0.02$, $V-R=-0.01\pm0.02$ and
  $R-I=0.16\pm0.04$. The mean colour and standard deviation of all
  stars in the 1A~1246$-$588 region is $B-V=1.03\pm0.22$ and
  $R-I=0.73\pm0.28$, and both star~X and A are significantly bluer
  than this. Furthermore, at the observed $R$-band magnitude and $R-I$
  colour, both stars are about half a magnitude bluer than the bulk of
  the stars at the same $R$-band magnitude. The optical position of
  star~X is
  $\alpha_\mathrm{J2000}=12^\mathrm{h}49^\mathrm{m}39\fs364\pm0\farcs06$,
  $\delta_\mathrm{J2000}=-59\degr05\arcmin14\farcs68\pm0\farcs05$,
  which is only $2\farcs4$ from the \emph{ROSAT} position and well
  within the $1\sigma$ uncertainty of $6\arcsec$. Star~A on the other
  hand has
  $\alpha_\mathrm{J2000}=12^\mathrm{h}49^\mathrm{m}35\fs660\pm0\farcs06$,
  $\delta_\mathrm{J2000}=-59\degr05\arcmin12\farcs94\pm0\farcs05$,
  which is $30\arcsec$ (about $5\sigma$) from the \emph{ROSAT}
  position. We estimate that the probability of finding a star as blue
  as star X within the 95\% confidence error circle of 1A~1246$-$588
  is about 2\%. Hence, we identify star~X as the optical counterpart
  to 1A~1246$-$588.

  A single star is present within the \emph{Chandra} error circle of
  4U~1812$-$12 at
  $\alpha_\mathrm{J2000}=18^\mathrm{h}15^\mathrm{m}06\fs155\pm0\farcs19$,
  $\delta_\mathrm{J2000}=-12\degr05\arcmin46\farcs70\pm0\farcs21$. This
  position is only $0\farcs5$ from the X-ray position of 4U~1812$-$12
  \citep{wpk+03}. From the photometry we obtain $R=22.15\pm0.02$ and
  $R-I=1.53\pm0.03$. This star is not as blue as the counterpart of
  1A~1246$-$588, however, in light of the larger absorbing column for
  4U~1812$-$12, this is not surprising. Still, the counterpart is
  bluer than the bulk of the stars in the 4U~1812$-$12 region, which
  have $R-I=1.88\pm0.23$. Furthermore, there are 9 stars within a
  radius of $5\arcsec$ from the \emph{Chandra} position of
  4U~1812$-$12, and all, except the candidate counterpart, are redder
  than $R-I=1.73$. Finally, we note that the probability of a chance
  coincidence of a star within the 95\% confidence error circle is
  about 0.15\%. We conclude that the star inside the error circle is
  the optical counterpart to 4U~1812$-$12.

  \section{Discussion}\label{sec:discussion}
  We have identified the optical companions to the low-mass X-ray
  binaries 1A~1246$-$588 and 4U~1812$-$12 based on their positional
  coincidence with the X-ray position and their colours. The
  counterpart to the first has $V=19.45$, $B-V=0.22$, while that of
  4U~1812$-$12 is somewhat fainter at $R=22.15$, $R-I=1.53$.

  Due to its position somewhat out of the Galactic plane, the hydrogen
  absorption column $N_\mathrm{H}$ towards 1A~1246$-$588 is moderate
  and suggests $A_V=1.7$ \citep{ps95}. This is smaller than the
  maximum absorption in this line-of-sight, which is predicted to
  reach $A_V=1.9$ around $d=7$\,kpc by the model of
  \citet{dcl03}. This limit constrains the absolute magnitude of the
  companion of 1A~1246$-$588 to $M_V\ga3.5$. Though this reasoning
  assumes that both the model by \citet{ps95} and \citet{dcl03} are
  correct, this distance is in agreement with estimates from
  \emph{BeppoSAX}/WFC and RXTE/ASM observations of photospheric radius
  expansion bursts of 1A~1246$-$588, which suggest a distance of
  5\,kpc (in't Zand et al.\ in prep.).

  For 4U~1812$-$12, the photospheric radius expansion bursts that have
  been observed by \citet{cbn+00} provide an estimate on the distance
  to this LMXB.  Assuming an Eddington peak luminosity of
  $L_\mathrm{X}=3.8\times10^{38}$\,erg\,s$^{-1}$ for the accretion of
  helium-rich material \citep{khz+03}, the distance is estimated at
  4.6\,kpc.  If hydrogen-rich material is accreted instead, the
  Eddington luminosity of $\sim\!2\times10^{38}$\,erg\,s$^{-1}$
  reduces the distance to 3.4\,kpc (see also
  \citealt{jn04}). Furthermore, the value of $N_\mathrm{H}$ derived
  from the X-ray absorption suggests $A_V=6.4$ \citep{ps95} and, using
  the relative extinction coefficients of \citet{sfd98}, $A_R=5.2$. As
  such, the optical companion of 4U~1812$-$12 has an absolute $R$-band
  magnitude in the range of 3.6--4.2. If we assume that the
  counterpart has the same intrinsic colours as the counterpart of
  1A~1246$-$588, which has $(V-R)_0=-0.2$, this would translate to
  $M_V=3.4$--4.0.

  These absolute magnitudes place both systems amongst the
  intrinsically fainter of the LMXBs known. According to
  \citet{vpm94}, this suggests that these systems are ultra-compact
  X-ray binaries (UCXBs; having an orbital period below an
  hour). Here, the faintness of the counterpart is due to the
  reprocession of X-rays in a physically small accretion
  disk. However, these systems remain candidate UCXBs until the
  orbital period is determined, i.e.\ either through optical/IR or
  X-ray observations.

  If these systems indeed turn out to have ultra-compact orbits, it is
  interesting to note that for the observed X-ray luminosities of
  $L_\mathrm{X}\approx0.9\times10^{36}$\,erg\,cm$^{-2}$\,s$^{-1}$
  (1--10\,keV) for 4U~1812$-$12 and
  $L_\mathrm{X}\la10^{36}$\,erg\,cm$^{-2}$\,s$^{-1}$ (0.1--2.4\,keV)
  for 1A~1246$-$588, these systems satisfy the notion presented by
  \citet{zcm05}; that LMXBs with persistent luminosities with
  $L_\mathrm{X}\la10^{36}$\,erg\,cm$^{-2}$\,s$^{-1}$ may be
  ultra-compact X-ray binaries.

  \begin{acknowledgements}
    The Munich Image Data Analysis System (MIDAS) is developed and
    maintained by the European Southern Observatory. This research
    made use of results provided by the ASM/RXTE teams at MIT and at
    the RXTE SOF and GOF at NASA's GSFC. CGB and JJMZ acknowledge
    support by the Netherlands Organization for Scientific Research
    (NWO). PGJ acknowledges funding from NASA grant GO4-5033X.
  \end{acknowledgements}

  \bibliographystyle{aa}

\end{document}